\documentclass[sigconf]{acmart}
\usepackage{booktabs} 
\usepackage{makecell}

\usepackage[T1]{fontenc}
\usepackage{epstopdf}
\usepackage{multirow}
\usepackage{balance}
\usepackage{amsfonts}
\usepackage{amsmath}
\usepackage{amssymb}
\usepackage{graphicx}
\usepackage{microtype}
\usepackage{url}
\usepackage{enumitem}
\usepackage{wrapfig,lipsum}
\usepackage{multirow}
\usepackage{makecell}
\usepackage{wrapfig}

\usepackage[boxed, ruled,vlined,linesnumbered]{algorithm2e}
\usepackage{algpseudocode}
\usepackage{microtype}

\usepackage{threeparttable}
\usepackage{subfigure}
\usepackage{booktabs}
\usepackage{caption}

\newcommand{\eg}{\textit{e.g.}}
\newcommand{\ie}{\textit{i.e.}}
\newcommand{\sepattn}{\textbf{\textsf{SepAttn}}\xspace}
\newcommand{\bs}[1]{\boldsymbol{#1}}

\usepackage[compact]{titlesec}
\titlespacing{\section}{0pt}{2ex}{1ex}
\titlespacing{\subsection}{0pt}{1ex}{0ex}
\titlespacing{\subsubsection}{0pt}{0.5ex}{0ex}

\AtBeginDocument{%
  \providecommand\BibTeX{{%
    \normalfont B\kern-0.5em{\scshape i\kern-0.25em b}\kern-0.8em\TeX}}}

\setcopyright{acmcopyright}

\copyrightyear{2020}
\acmYear{2020}
\setcopyright{rightsretained}
\acmConference[WSDM '20]{The Thirteenth ACM International Conference on Web Search and Data Mining}{February 3--7, 2020}{Houston, TX, USA}
\acmBooktitle{The Thirteenth ACM International Conference on Web Search and Data Mining (WSDM '20), February 3--7, 2020, Houston, TX, USA}
\acmDOI{10.1145/3336191.3371775}
\acmISBN{978-1-4503-6822-3/20/02}

\begin{document}
\fancyhead{}
\title{Separate and Attend in Personal Email Search}

\author{Yu Meng$^{1*}$, Maryam Karimzadehgan$^2$, Honglei Zhuang$^{2}$, Donald Metzler$^2$}
\affiliation{
\institution{$^1$Department of Computer Science, University of Illinois at Urbana-Champaign, IL, USA} 
\institution{$^2$Google LLC, Mountain View, CA, USA}
\institution{$^{1}$yumeng5@illinois.edu \ \ \ $^2$\{maryamk, hlz, metzler\}@google.com}
}

\thanks{$^*$Work done while Yu Meng was at Google.}
\renewcommand{\shortauthors}{Meng et al.}


\begin{abstract}
In personal email search, user queries often impose different requirements on different aspects of the retrieved emails. For example, the query ``my recent flight to the US'' requires emails to be ranked based on both textual contents and recency of the email documents, while other queries such as ``medical history'' do not impose any constraints on the recency of the email.  Recent deep learning-to-rank models for personal email search often directly concatenate dense numerical features\footnote{Throughout the paper, we use dense features to refer to dense numerical features that cannot be derived from sparse features; not to be confused with embedded sparse features.} (\eg, document age) with embedded sparse features (\eg, n-gram embeddings). In this paper, we first show with a set of experiments on synthetic datasets that direct concatenation of dense and sparse features does not lead to the optimal search performance of deep neural ranking models. 
To effectively incorporate both sparse and dense email features into personal email search ranking, we propose a novel neural model, \sepattn. \sepattn first builds two separate neural models to learn from sparse and dense features respectively, and then applies an attention mechanism at the prediction level to derive the final prediction from these two models. 
We conduct a comprehensive set of experiments on a large-scale email search dataset, and demonstrate that our \sepattn model consistently improves the search quality over the baseline models.
\end{abstract}


\keywords{Learning-to-Rank, Email Search, Neural Attention Model}

\maketitle


\section{Introduction}
Email has long been an important means of daily communication. Personal email search, which helps users to quickly retrieve the emails they are looking for from their own corpora, has been an intriguing research topic in information retrieval (IR) for years. Email search is formulated as a learning-to-rank problem, which has been tackled with different learning models, such as boosted trees~\cite{Burges2010FromRT}, SVM-based linear models~\cite{Joachims2002OptimizingSE,Cao2006AdaptingRS,CarmelCIKM2015}, and shallow neural networks~\cite{Burges2005LearningTR,Cao2007LearningTR}.

Recently, deep neural networks (DNNs) have shown great success in learning-to-rank tasks. They significantly improve the performance of search engines in the presence of large-scale query logs in both web search~\cite{GuoCIKM2016} and email settings~\cite{TranSIGIR2019,ShenCIKM2018,ZamaniWWW2017}. The advantages of DNNs over traditional models are mainly two-fold: (1) DNNs have strong power to learn embedded representations from sparse features, including words~\cite{Mikolov2013DistributedRO} and characters~\cite{Bojanowski2016EnrichingWV}. This allows effective and accurate matching of textual features between queries and documents. (2) DNNs are proved to have universal approximation capability~\cite{Hornik1991ApproximationCO} and thus are able to capture high-order interactions between query and document features.


In the personal email search scenario, user queries impose different requirements on different aspects of email documents to be retrieved. For example, the query ``my recent flight to the US'' requires the email search system to focus on both the textual contents and the recency of email documents, while queries such as ``medical history'' expect emails to be retrieved regardless of the recency. In email search models, different properties of email documents are reflected by different types of features, including dense numerical ones (\eg, document age) and sparse categorical ones (\eg, n-grams). However, there have been few efforts that study how to effectively exploit both dense and sparse features in the learning-to-rank setting, probably because a natural approach exists---simply concatenating dense features with embedded sparse features and feeding them into the DNNs. Indeed, many previous deep neural email search models use direct concatenation of dense features with embedded sparse features~\cite{Dehghani2017NeuralRM,Severyn2015LearningTR,ShenCIKM2018,TranSIGIR2019}.

In this paper, we begin with a set of empirical findings and analyses to show that direct concatenation of dense features with embedded sparse features does not lead to the optimal performance of DNN-based ranking models. As a result, we propose the \sepattn model as an effective way to incorporate both dense and sparse features into the personal email search model.
More specifically, \sepattn model consists of three major modules: \textbf{(1)} Separate DNN models to learn from sparse and dense features, respectively. \textbf{(2)} An attention mechanism that aggregates the outputs from the sparse feature and dense feature DNN models. \textbf{(3)} A regularizer that enables joint learning of the sparse and dense feature DNN models. The main advantage of \sepattn over the simple concatenation approach is that \sepattn separates dense features from sparse ones, and automatically learns to \emph{explicitly} focus on the important feature set, while ignore the unimportant ones for different query types.


In summary, the followings are the contributions of this paper:
\begin{itemize}
\item We empirically show that simply concatenating embedded sparse features with dense numerical features is sub-optimal for DNN-based neural ranking models.
\item To effectively leverage both sparse and dense features in neural ranking models, we propose to learn two separate models where their outputs are aggregated via an \textbf{attention} mechanism to derive the final output.
\item We propose a regularization method to train sparse and dense feature models in a collaborative manner.
\item We conduct a comprehensive set of experiments on synthetic datasets and a large-scale real-world email search dataset to demonstrate the advantage of our proposed method.
\end{itemize}

The rest of the paper is organized as follows. Section~\ref{sec:related} reviews related work. Section~\ref{sec:motiv} motivates the problem through a set of experiments on synthetic datasets. Section~\ref{sec:methodology} introduces our proposed \sepattn model. Section~\ref{sec:methodSyn} presents and analyzes the performances of \sepattn on the synthetic datasets. In Section~\ref{sec:main-exp}, we report and discuss the experimental results on a real-world large-scale email search dataset. Section~\ref{sec:conclude} concludes the paper and discusses future directions.


\section{Related Work}
\label{sec:related}
In this section, we review related works on learning-to-rank, email search models and state-of-the-art neural attention models.

\subsection{Learning-to-Rank}
Learning-to-rank refers to building ranking models with machine learning algorithms. In early years, learning-to-rank has been studied with different models, such as boosted trees~\cite{Burges2010FromRT}, SVM-based linear models~\cite{Joachims2002OptimizingSE,Cao2006AdaptingRS,CarmelCIKM2015} and shallow neural networks~\cite{Burges2005LearningTR,Cao2007LearningTR}. Recent years have witnessed great success of applying DNNs to learning-to-rank, such as~\cite{Borisov2016ANC,Dehghani2017NeuralRM,Pang2017DeepRankAN,Severyn2015LearningTR}. For a complete literature review on neural ranking models for information retrieval, please refer to a survey by Mitra and Craswell~\cite{Mitra2017NeuralMF}.

\subsection{Email Search}
There have been several studies in the IR community focusing on the task of email search.  
The Enterprise tracks of TREC 2005~\cite{SoboroffTREC2005} and TREC 2006~\cite{SoboroffTREC2006} provide public datasets containing email data and summarize some early explorations~\cite{OgilvieTREC2005,CraswellTREC2005,MacdonaldSIGIR2006}. 
A typical trade-off in email search system is to balance the importance of content-based relevance and other features, \eg~freshness.  
Carmel et al.~\cite{CarmelCIKM2015} proposed an email search framework with a learning-to-rank re-ranking module that combines freshness with relevance signals of emails as well as other features such as user actions.  
Alternatively, Carmel et al.~\cite{CarmelWWW2017} studied to present users with both the relevance-ranked results as well as the time-ranked results in two separate lists for better user experience.  
A number of studies specifically focus on improving the content-based relevance signals in email search.  
Kuzi et al.~\cite{KuziSIGIR2017} explored several methods to expand the usually short and sparse queries by finding more related terms to improve the relevance results. 
Li et al.~\cite{LiSIGIR2019} studied a more specific synonym expansion problem to improve email search performance.

User interaction data such as clicks is another important signal for learning-to-rank models in email search.
Bendersky et al.~\cite{BenderskyWSDM2017} leveraged user interactions by attribute parameterization.  
Wang et al.~\cite{WangWSDM2018} mitigated the position bias in click data for better training of the model.
In addition, Zamani et al.~\cite{ZamaniWWW2017} showed that contexts such as search request time and location of users were helpful for email search quality.

There are also studies on understanding and leveraging query intent information in email search.  Ai et al.~\cite{AiWWW2017} conducted a thorough survey of search intent by analyzing user logs of email search.  Shen et al.~\cite{ShenCIKM2018} categorized email search queries into different clusters before adding the query cluster information to improve email ranking.  

To the best of our knowledge, there is no previous work on email search that studied how to effectively combine dense numerical features with embedded sparse features in DNN-based ranking models. In the previous works, dense features were directly concatenated with embedded sparse features, which led to the sub-optimal performances of DNN ranking models, as we will show later.

\subsection{Attention Models}
Recently, neural attention mechanisms have demonstrated enormous power on sequence modeling. They derived the optimal sequence representation by learning to focus on the important tokens in the sequence and down-weighting unimportant ones for downstream tasks. The attention mechanism was first proposed by Bahdanau et al.~\cite{Bahdanau2015NeuralMT} in machine translation, where attention was used on top of RNN encoders for input-output alignment. Later, the attention mechanism has been adapted to a wide range of compelling sequence modeling tasks, including image caption generation~\cite{Xu2015ShowAA}, text classification~\cite{Yang2016HierarchicalAN} and natural language question answering~\cite{Hermann2015TeachingMT,Kumar2015AskMA,Sukhbaatar2015EndToEndMN}. 

The above studies employ attention mechanism in conjunction with RNN or CNN models. Vaswani et al.~\cite{Vaswani2017AttentionIA} proposed the Transformer, which used self-attention along with positional encoding. Later, Devlin et al.~\cite{Devlin2019BERTPO} proposed a deep bidirectional Transformer structure, BERT, which becomes one of the state-of-the-art pre-trained language models benefiting many downstream tasks with fine-tuning. The attention mechanism has also been generalized to attend to a group of structurally adjacent items instead of single ones, as studied by Li et al.~\cite{li2019area}. 

Regardless of the manner that attention mechanisms are used in the previous works, the common purpose was to derive sequence representations as a weighted average of token representations. Therefore, the attention mechanisms were applied at the \emph{feature level}, as a step of feature learning. In our work, however, the attention is used to aggregate the outputs from two models---the sparse and the dense feature models, to derive the final scoring outputs of the ranking model. To the best of our knowledge, this is the first work that applies attention mechanism at the \emph{prediction level}.


\begin{figure*}[h]
\centering
\subfigure[First Synthetic dataset (Sparse only)]{
\includegraphics[width=0.3203\linewidth]{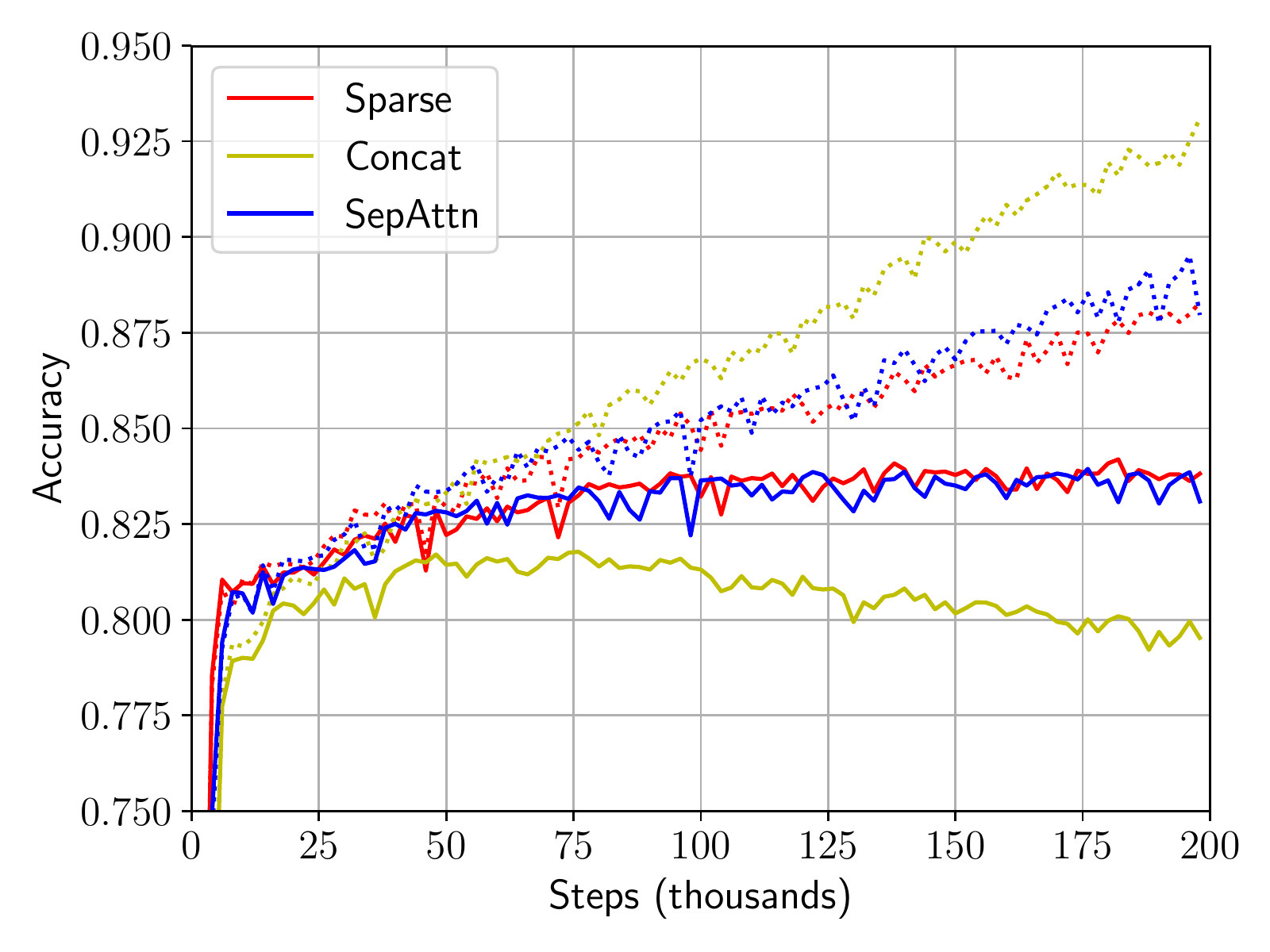}
\label{fig:sparse}
}
\subfigure[Second Synthetic dataset (Dense only)]{
\includegraphics[width=0.3203\linewidth]{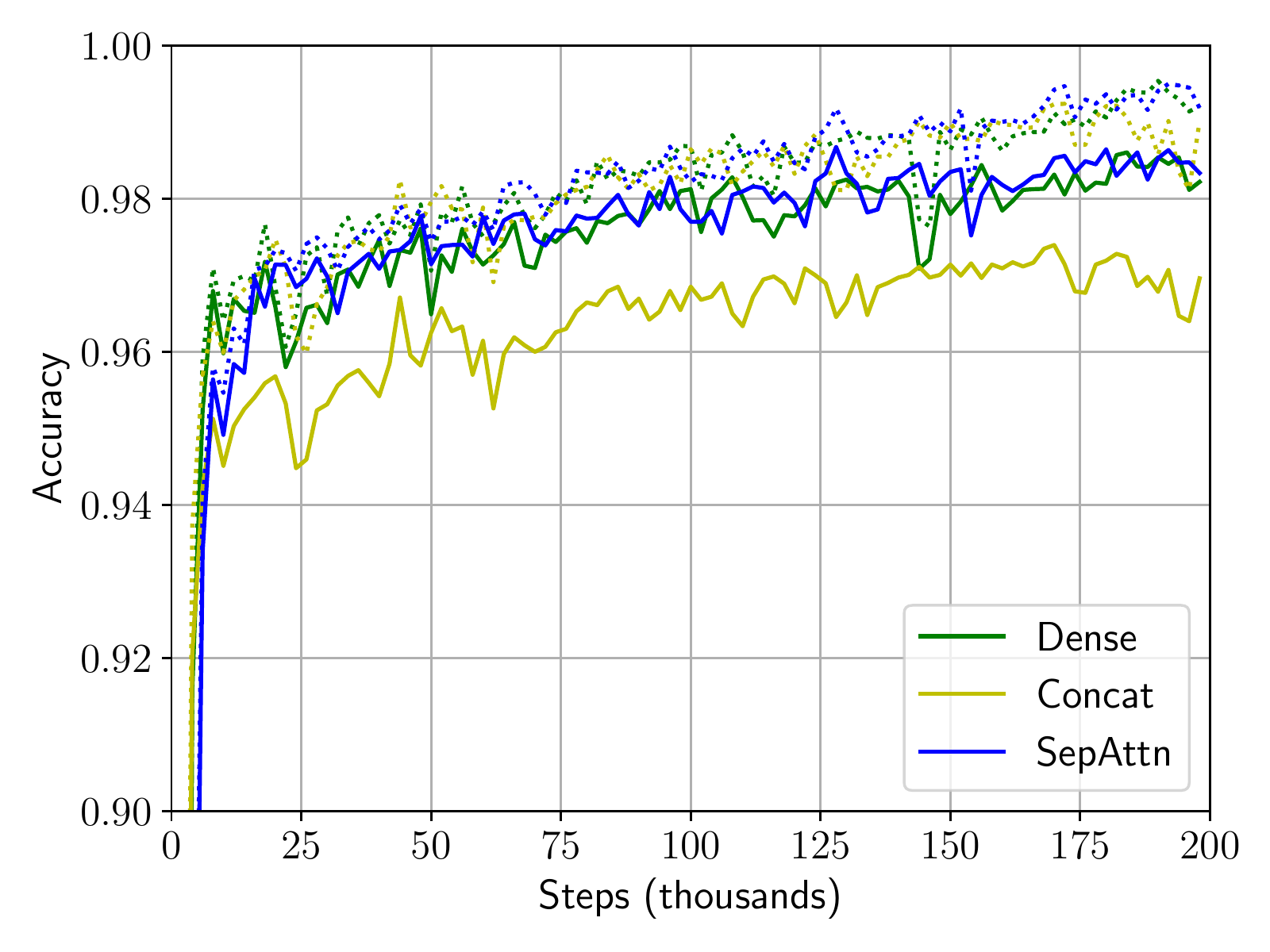}
\label{fig:dense}
}
\subfigure[Third Synthetic dataset (Sparse and Dense Combined)]{
\includegraphics[width=0.3203\linewidth]{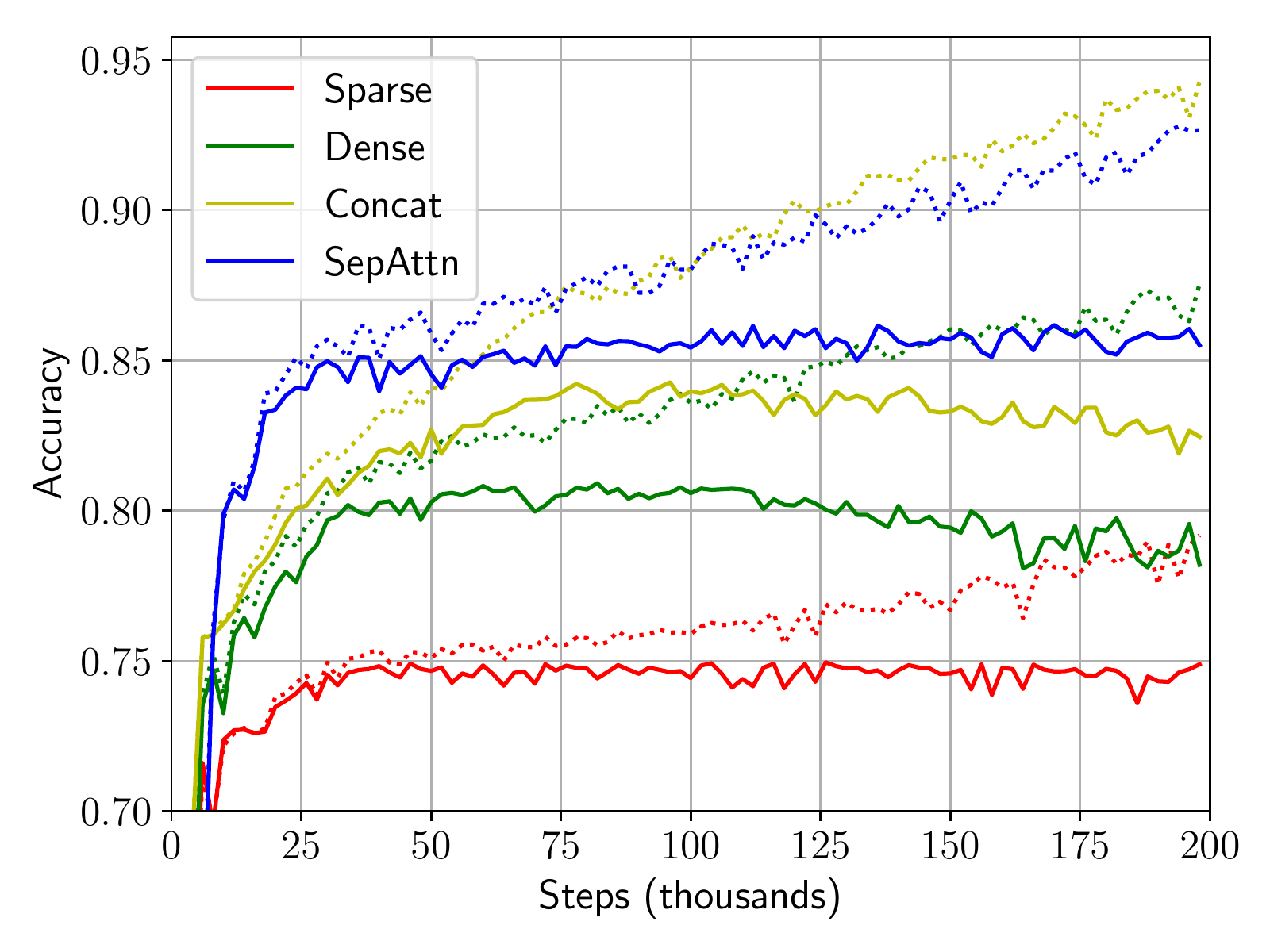}
\label{fig:and}
}
\caption{Training and testing performance of three models, \ie, Sparse only, Dense only and Sparse + Dense models on three synthetic datasets. The dotted lines represent training accuracy and the solid lines represent testing accuracy. For completeness, we also show the performance of our \sepattn model in these figures. ``Accuracy'' is the ratio of correctly classified documents.}
\label{fig:syn}
\end{figure*}

\section{Motivation}
\label{sec:motiv}

In this section, we provide motivation for our proposed ``Separate and Attend'' model, \ie, \textbf{\sepattn}, and show that simply concatenating different feature types (\ie, dense and sparse features) does not lead to the optimal performance. Different queries in email search impose different requirements. For example, a query such as ``medical history'' does not indicate requirements on dense features (\eg, recency of an email document). As a result, simply concatenating both dense and sparse features and feeding them into the model might lead to sub-optimal performance for such queries. In such a case, dense features are purely noise and could potentially distract the model from concentrating on useful features.  

In order to simulate the personal email search scenario, we create three synthetic datasets and consider a simple binary classification problem, where a document $d$ is classified to be relevant ($y=1$) or irrelevant ($y=0$) for a given query $q$. We use $100$-dimensional pre-trained Joint Spherical Embeddings (JoSE)~\cite{Meng2019SphericalTE} as embedded sparse feature representations for queries and documents.

The query features $\bs{q}$ and document sparse features $\bs{d}_{\text{sparse}}$ are obtained by randomly sampling word embeddings from the pre-trained embedding vocabulary. The document dense features $\bs{d}_{\text{dense}}$ are generated from a uniform distribution in the interval $[-0.5, 0.5]$. We set the dimension of $\bs{d}_{\text{sparse}}$ (after embedding) and $\bs{d}_{\text{dense}}$ to be both $100$. The ground-truth labels are generated differently for each of these synthetic datasets which will be explained later in this section.

We compare the performance of these models: \textbf{(1)} The model that learns from $\bs{q}$ and $\bs{d}_{\text{sparse}}$ only; \textbf{(2)} The model that learns from $\bs{q}$ and $\bs{d}_{\text{dense}}$ only; \textbf{(3)} The model that learns from $\bs{q}$ and the concatenation of $\bs{d}_{\text{sparse}}$ and $\bs{d}_{\text{dense}}$.  All models are built based on a feed-forward DNN~\cite{JeffDNN} with a hidden layer of $50$ dimensions; the only difference between the models is that they learn from different feature sets. We generate $20,000$ samples for both training and testing sets.

For the \textbf{first} dataset, the following rule describes how the ground-truth data is generated:

$$
y = \begin{cases}
1, & \text{if} \, \cos(\bs{d}_{\text{sparse}}, \bs{q}) > 0\\
0, & \text{else}
\end{cases}
$$
which simulates the scenario where document and query matching is based only on sparse features, and dense features are unimportant/unused.

Similarly, we create the \textbf{second} synthetic dataset by generating ground-truth labels only based on dense features as below:
$$
y = \begin{cases}
1, & \text{if}\, \bs{q}\in \mathcal{Q}\, \text{and}\, \sum \bs{d}_{\text{dense}} < 0\\
0, & \text{else}
\end{cases}
$$
where $\mathcal{Q}$ is a manually selected vocabulary of words that impose requirements on the recency of the documents, such as ``recent'', ``latest'', ``newest'', ``today''.

We show the performance of the aforementioned models on the first synthetic dataset and second synthetic dataset in Figure~\ref{fig:sparse} and Figure~\ref{fig:dense} respectively. In these figures, dotted lines represent training curves, and solid lines represent testing curves. We also show the performance of our proposed method \textbf{\sepattn} for completeness in these figures and postpone the discussions about its performance to Section~\ref{sec:methodSyn}.

On the first synthetic dataset, sparse only model is the benchmark because it learns only from sparse features and inherently avoids unimportant dense features. We observe from Figure~\ref{fig:sparse} that the concatenation model overfits the training set---when its training accuracy goes up, its testing accuracy goes down. This indicates that the concatenation model learns noisy signals from the training set that cannot be generalized to the testing set, demonstrating its ineffectiveness of discriminating the useful features from unimportant ones.

On the second synthetic dataset, dense only model is the benchmark since it avoids the unimportant sparse features. Figure~\ref{fig:dense} shows that the testing accuracy of concatenation model falls behind that of the dense only model. This again demonstrates that the concatenation model is negatively influenced by unimportant features.

The \textbf{third} synthetic dataset is generated based on combination of both sparse and dense features and the ground-truth labels are defined as follows:

$$
y = \begin{cases}
1, & \text{if} \, \big(\cos(\bs{d}_{\text{sparse}}, \bs{q}) > 0 \big) \land \big(\bs{q}\in \mathcal{Q}\, \text{and}\, \sum \bs{d}_{\text{dense}} < 0 \big)\\
0, & \text{else}
\end{cases}
$$
where the document-query matching requires both sparse and dense feature matching.

The performance of the above models using this dataset is shown in Figure~\ref{fig:and}. Although the concatenation model outperforms sparse only and dense only models, it still suffers from overfitting (testing accuracy increases while training accuracy decreases). This demonstrates that DNNs are ineffective to learn from concatenated sparse and dense features. We explain the reason as below: $\bs{d}_{\text{dense}}$ and $\bs{d}_{\text{sparse}}$ come from different feature space---dense numerical feature values (\eg, document age) have practical meanings (\ie, a higher value of document age means the email was received earlier), while embedding feature values (\eg, n-gram embedding) do not have practical meanings~\cite{Subramanian2017SPINESI} (\ie, a higher value in n-gram embedding dimensions is not interpretable). If $\bs{d}_{\text{dense}}$ and $\bs{d}_{\text{sparse}}$ are directly concatenated and fed into the DNNs, operations applied between $\bs{d}_{\text{dense}}$ and $\bs{d}_{\text{sparse}}$ are meaningless (\eg, it does not make sense to add/multiply document age with n-gram embeddings) and can lead to ineffectiveness and overfitting of DNNs. 

In the next section, we describe the details of our model, \ie, \textbf{\sepattn}.


\section{Methodology}
\label{sec:methodology}
We first formulate the problem and define the notations. Then we describe our \sepattn model. 
\subsection{Problem Formulation}

The inputs for personal email search problem are tuples $(q, \mathcal{D})$ where $q$ is a user query string, and $\mathcal{D}=\{d_1,\dots,d_n\}$ is a list of candidate documents. The ground truth labels $\bs{y} \in \{0,1\}^n$ are user click-through data indicating whether the corresponding document is clicked. The goal of personal email search model is to rank $\mathcal{D}$ so that the clicked document is ranked as high as possible.


The query features $\bs{q}$ are n-grams and character n-grams extracted from the original user query string $q$, the document features $\bs{d}$ include sparse categorical features $\bs{d}_{\text{sparse}}$ (\eg, n-grams) and dense numerical features $\bs{d}_{\text{dense}}$ (\eg, document ages).

To preserve privacy, the query-document inputs are anonymized based on $k$-anonymity approach~\cite{Sweeney2002kAnonymityAM}, and only query and document n-grams that are frequent in the entire corpus are retained. We summarize the features used in the model in Table~\ref{tab:features}.

\begin{table}[h]
  \caption{Features used in personal email search.}
  \label{tab:features}
  \scalebox{1}{
  \begin{tabular}{cc}
    \toprule
    Type & Features\\
    \midrule
    \multirow{2}{*}{Query Feature ($\bs{q}$)} & n-grams \\ 
    & character n-grams\\
    \midrule
    \multirow{2}{*}{Document Sparse Feature ($\bs{d}_{\text{sparse}}$)} & n-grams \\
    & character n-grams \\
    \midrule
    \multirow{3}{*}{Document Dense Feature ($\bs{d}_{\text{dense}}$)} & number of attachments\\
    & number of recipients\\
    & document age\\
  \bottomrule
\end{tabular}
}
\end{table}

Given the input features $\mathcal{X} = \{\bs{q}, \bs{d}_1, \dots, \bs{d}_n\}$, the email search model learns a per-item scoring function $h: (\bs{q}, \bs{d}_i) \to \mathbb{R}$, such that
\begin{equation}
\label{eq:func}
\bs{h}(\mathcal{X}) = \begin{bmatrix}
h(\bs{q}, \bs{d}_1) & \cdots & h(\bs{q}, \bs{d}_n)
\end{bmatrix}^\top
\end{equation}
induces a permutation $\pi$ and $\bs{h}(\mathcal{X})\big\rvert_{\pi^{-1}(r)}$ monotonically decreases with increasing rank $r$. The per-item scoring function $h$ could be learned by gradient boosted trees~\cite{Friedman2001GreedyFA}, support vector machines~\cite{Joachims2006TrainingLS} or neural networks~\cite{Burges2005LearningTR}. In this paper, we aim to design a DNN-based model that effectively incorporates both $\bs{d}_{\text{sparse}}$ and $\bs{d}_{\text{dense}}$ to improve personal email search performance.

With the input features $(\bs{q}, \bs{d}_i)$ described above, ground truth click-through data $y_i$ and a per-item scoring function $h$, we follow the Softmax Cross-Entropy listwise loss setting in~\cite{Pasumarthi2019TFRankingST} to train the model. Specifically, the loss is defined as
\begin{equation}
\label{eq:softmax}
\mathcal{L}_{\text{listwise}} = - \sum_{i=1}^{n} y_i \log \left( \frac{\exp\left({h(\bs{q}, \bs{d}_i)}\right)}{\sum_{j=1}^n \exp\left({h(\bs{q}, \bs{d}_j)}\right)} \right).
\end{equation}



\subsection{Separate and Attend Method}
\label{sec:sepattn}
\begin{figure*}[h]
  \centering
  \includegraphics[width=\linewidth]{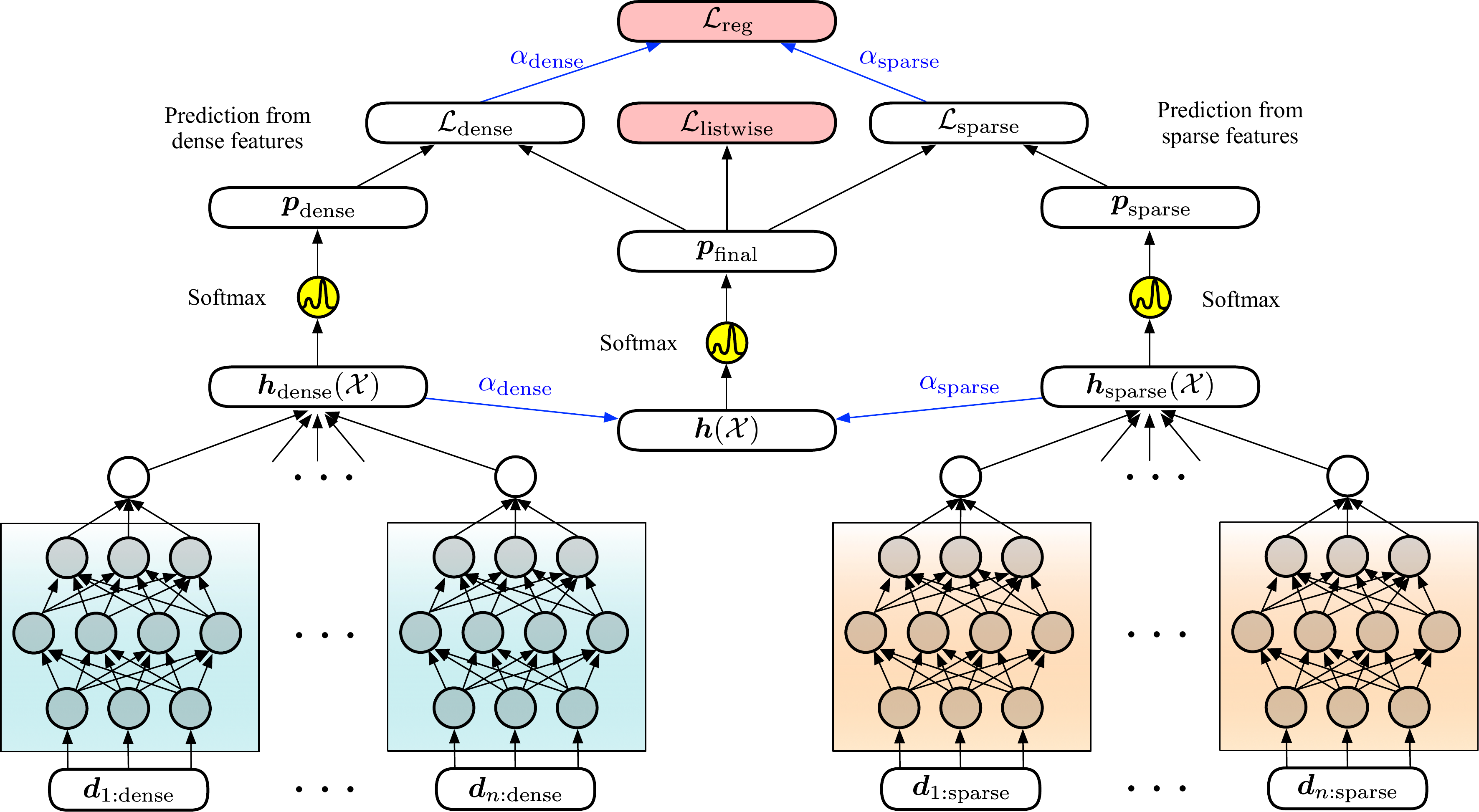}
  \caption{The overview of \sepattn model. \sepattn builds separate models to learn from sparse and dense features, and then aggregates the two models' outputs via an attention mechanism. \sepattn also employs a regularization term to enable joint training of the two models, and the final loss function is a combination of the listwise loss with the regularization loss. For simplicity, we omit the query features which are concatenated with the document dense/sparse features before fed into the DNN models.}
  \label{fig:model}
\end{figure*}

In this section, we introduce our ``Separate and Attend model'', \ie, \textbf{\sepattn} model, to address the observed DNN learning issues when sparse and dense features are directly concatenated, as described in Section~\ref{sec:motiv}. The model structure of \sepattn is presented in Figure~\ref{fig:model}.

Often times different queries impose different requirements regarding an email document to be retrieved, and different email document properties are reflected by the sparse and dense features extracted. In this section, we first build separate models that learn from each sparse and dense features individually, and then develop an \textbf{attention} mechanism at the prediction stage that enables the model to explicitly learn to focus on the important features for different queries. 

Specifically, we first build two models to learn two per-item scoring functions (one for sparse features and the other one for dense features):
$$
h_{\text{sparse}}: (\bs{q}, \bs{d}_{i:\text{sparse}}) \to \mathbb{R}, \quad h_{\text{dense}}: (\bs{q}, \bs{d}_{i:\text{dense}}) \to \mathbb{R},
$$
which score a document given a query based on sparse and dense document features, respectively.

Then the scores of a list of $n$ documents given by sparse and dense feature models are:
\begin{equation*}
\bs{h}_{\text{sparse}}(\mathcal{X}) = \begin{bmatrix}
h_{\text{sparse}}(\bs{q}, \bs{d}_{1:\text{sparse}}) & \cdots & h_{\text{sparse}}(\bs{q}, \bs{d}_{n:\text{sparse}})
\end{bmatrix}^\top,
\end{equation*}
\begin{equation*}
\bs{h}_{\text{dense}}(\mathcal{X}) = \begin{bmatrix}
h_{\text{dense}}(\bs{q}, \bs{d}_{1:\text{dense}}) & \cdots & h_{\text{dense}}(\bs{q}, \bs{d}_{n:\text{dense}})
\end{bmatrix}^\top.
\end{equation*}

Next, we introduce an \textbf{attention} mechanism that aggregates $\bs{h}_{\text{sparse}}(\mathcal{X})$ and $\bs{h}_{\text{dense}}(\mathcal{X})$ to derive the final scores for the document list. Specifically, we first feed $\bs{h}_{\text{sparse}}(\mathcal{X})$ and $\bs{h}_{\text{dense}}(\mathcal{X})$ to a one-layer feed-forward neural network~\cite{JeffDNN} to derive the hidden representation of the document scores, and then compute their attention weights based on the hidden representations as well as a context vector.
Finally, $\bs{h}_{\text{sparse}}(\mathcal{X})$ and $\bs{h}_{\text{dense}}(\mathcal{X})$ are weighted-averaged according to their attention weights to derive the final scores $\bs{h}(\mathcal{X})$ for all the documents. Mathematically, the \textbf{attention} mechanism is presented as:
\begin{align}
\bs{u}_{\text{sparse}} &= \tanh(W \bs{h}_{\text{sparse}}(\mathcal{X}) + \bs{b}), \nonumber\\
\bs{u}_{\text{dense}} &= \tanh(W \bs{h}_{\text{dense}}(\mathcal{X}) + \bs{b}), \nonumber\\
\alpha_{\text{sparse}} &= \frac{\exp(\bs{u}_{\text{sparse}}^\top \bs{v})}{\exp(\bs{u}_{\text{sparse}}^\top \bs{v}) + \exp(\bs{u}_{\text{dense}}^\top \bs{v})}, \label{eq:alpha_s} \\
\alpha_{\text{dense}} &= \frac{\exp(\bs{u}_{\text{dense}}^\top \bs{v})}{\exp(\bs{u}_{\text{sparse}}^\top \bs{v}) + \exp(\bs{u}_{\text{dense}}^\top \bs{v})}, \label{eq:alpha_d}\\
\bs{h}(\mathcal{X}) &= \alpha_{\text{sparse}} \bs{h}_{\text{sparse}}(\mathcal{X}) + \alpha_{\text{dense}} \bs{h}_{\text{dense}}(\mathcal{X}), \label{eq:final}
\end{align}
where $W \in \mathbb{R}^{n \times n}$ and $\bs{b} \in \mathbb{R}^{n}$ are trainable weights and bias vector respectively; $\bs{v} \in \mathbb{R}^{n}$ is the context vector, randomly initialized and trained together with other parameters.

Here $\alpha_{\text{sparse}}$ and $\alpha_{\text{dense}}$ are \textbf{attention weights} assigned to the sparse feature model and the dense feature model, respectively. As shown in Equation~(\ref{eq:final}), the higher the attention weight a model gets, the more dominant role it plays in the final result.

We explain why the above attention mechanism enables the model to focus on important features for different queries. The hidden representations $\bs{u}_{\text{sparse}}$ and $\bs{u}_{\text{dense}}$ can be interpreted as encoding the importance of $\bs{h}_{\text{sparse}}(\mathcal{X})$ and $\bs{h}_{\text{dense}}(\mathcal{X})$ respectively. For example, if dense features are noisy features (\ie, not important for a given query which might mislead the model) with respect to a specific query, $\bs{h}_{\text{dense}}(\mathcal{X})$ will be random across all documents, and $\bs{u}_{\text{dense}}$ will encode that this is an unimportant result. The context vector $\bs{v}$ can be seen as a representation of important results, and therefore a hidden representation similar to $\bs{v}$ will be assigned higher weights. In this way, the unimportant/noisy features will be down-weighted by the attention mechanism, and predictions from the important features will dominate the final score.

In the following section, we describe the model learning with regularization. 

\subsection{Joint Learning with Regularization}
Until now, the dense and sparse feature models are trained without any interactions, \ie, the outputs of dense feature model will not directly influence the training of the sparse feature model, and vice versa. To enable direct interactions between the two models, we introduce a regularization term into \textbf{\sepattn}, motivated by the idea of co-training~\cite{Blum1998CombiningLA} that when two learning models capture different and complementary feature sets of the same instances, the two models can mutually enhance each other during training. 

To encourage the dense and sparse feature models to output consistent results given the same set of documents, a straightforward way is to minimize the difference between $\bs{h}_{\text{sparse}}(\mathcal{X})$ and $\bs{h}_{\text{dense}}(\mathcal{X})$. However, since the two sets of scores are computed from different feature sets, they will be in different scales, making it ineffective to regularize directly on $\bs{h}_{\text{sparse}}(\mathcal{X})$ and $\bs{h}_{\text{dense}}(\mathcal{X})$. Therefore, we instead regularize on the probability distributions after normalizing $\bs{h}_{\text{sparse}}(\mathcal{X})$ and $\bs{h}_{\text{dense}}(\mathcal{X})$ with Softmax, \ie,
{\small
$$
\bs{p}_{\text{sparse}} = \begin{bmatrix} \frac{\exp\left({h_{\text{sparse}}(\bs{q}, \bs{d}_{1:\text{sparse}})}\right)}{\sum_{j=1}^n \exp\left({h_{\text{sparse}}(\bs{q}, \bs{d}_{j:\text{sparse}})}\right)} & \cdots & \frac{\exp\left({h_{\text{sparse}}(\bs{q}, \bs{d}_{n:\text{sparse}})}\right)}{\sum_{j=1}^n \exp\left({h_{\text{sparse}}(\bs{q}, \bs{d}_{j:\text{sparse}})}\right)} \end{bmatrix}^\top,
$$
$$
\bs{p}_{\text{dense}} = \begin{bmatrix} \frac{\exp\left({h_{\text{dense}}(\bs{q}, \bs{d}_{1:\text{dense}})}\right)}{\sum_{j=1}^n \exp\left({h_{\text{dense}}(\bs{q}, \bs{d}_{j:\text{dense}})}\right)} & \cdots & \frac{\exp\left({h_{\text{dense}}(\bs{q}, \bs{d}_{n:\text{dense}})}\right)}{\sum_{j=1}^n \exp\left({h_{\text{dense}}(\bs{q}, \bs{d}_{j:\text{dense}})}\right)} \end{bmatrix}^\top.
$$} 

We then minimize the KL divergence of $\bs{p}_{\text{sparse}}$ and $\bs{p}_{\text{dense}}$ with respect to the final prediction $\bs{p}_{\text{final}}$:
\begin{align*}
\mathcal{L}_{\text{sparse}} &= D_{\text{KL}}(\bs{p}_{\text{final}} \;\|\; \bs{p}_{\text{sparse}}) = \sum_{i=1}^n p_{i:\text{final}} \log\left( \frac{p_{i:\text{final}}}{p_{i:\text{sparse}}} \right),\\
\mathcal{L}_{\text{dense}} &= D_{\text{KL}}(\bs{p}_{\text{final}} \;\|\; \bs{p}_{\text{dense}}) = \sum_{i=1}^n p_{i:\text{final}} \log\left( \frac{p_{i:\text{final}}}{p_{i:\text{dense}}} \right),
\end{align*}
where $\bs{p}_{\text{final}}$ is \sepattn's final prediction after normalizing $\bs{h}(\mathcal{X})$ in Equation~(\ref{eq:final}) via Softmax.

The regularization loss becomes the weighted average of $\mathcal{L}_{\text{sparse}}$ and $\mathcal{L}_{\text{dense}}$:
\begin{equation}
\mathcal{L}_{\text{reg}} = \alpha_{\text{sparse}} \mathcal{L}_{\text{sparse}} + \alpha_{\text{dense}} \mathcal{L}_{\text{dense}},
\end{equation}
where $\alpha_{\text{sparse}}$ and $\alpha_{\text{dense}}$ directly come from Equations~(\ref{eq:alpha_s}) and (\ref{eq:alpha_d}), respectively, and are automatically learned during training.

The purpose of including the attention weights $\alpha_{\text{sparse}}$ and $\alpha_{\text{dense}}$ again in the regularization term is to prevent noisy/unimportant features from misleading the regularization and disturbing the predictions learned from important features. For example, when dense features are noisy features with respect to a specific query, $\bs{p}_{\text{dense}}$ is a noisy distribution which is not meaningful to be regularized. Since the attention mechanism automatically down-weights unimportant features, the regularization term will encourage the interaction between the two models only when both feature sets are important.

Finally, the \textbf{\sepattn} model is trained via a combination of the listwise loss (Equation~(\ref{eq:softmax})) with the regularization loss:
\begin{equation}
\label{eq:tot}
\mathcal{L} = \mathcal{L}_{\text{listwise}} + \lambda \mathcal{L}_{\text{reg}},
\end{equation}
where $\lambda > 0$ is a hyperparameter that controls the importance of regularization. We will study its effect in model training in the experiment section.


\section{Results on Synthetic Datasets}
\label{sec:methodSyn}
In this section, we describe the results of our method on synthetic datasets described in Section ~\ref{sec:motiv}. From Figures~\ref{fig:sparse} and \ref{fig:dense}, we observe that \textbf{\sepattn} achieves comparable performance on the testing set to Sparse-only (benchmark on the first synthetic dataset) and Dense-only (benchmark on the second synthetic dataset) models, respectively, which shows the robustness of our model against noisy/unimportant features. To understand how \textbf{\sepattn} is able to focus on important features and ignore noisy ones, we visualize the attention weights, \ie, $\alpha_{\text{sparse}}$ and $\alpha_{\text{dense}}$ in Equations~(\ref{eq:alpha_s}) and (\ref{eq:alpha_d}) on the first and second synthetic datasets. We randomly select $20$ samples. It can be observed from Figure~\ref{fig:attn}(a) that on the first synthetic dataset, \textbf{\sepattn} assigns almost zero attention weights to the dense features, and effectively focuses on the sparse features. This explains why \textbf{\sepattn} can achieve similar performance with the Sparse-only model. Similarly, as shown in Figure~\ref{fig:attn}(b), \textbf{\sepattn} focuses mainly on the dense features for the second synthetic dataset, and does not pay attention to the sparse features.

In Figure~\ref{fig:and}, \textbf{\sepattn} performs the best on the testing set, demonstrating its effectiveness on learning from the combination of sparse and dense features. We visualize the attention weights in Figure~\ref{fig:attn}(c), where we use $0$/$1$ values to denote the classification outputs from sparse and dense feature models. For example, the first sample has a $0$ in the sparse row, and a $1$ in the dense row. This means that for the first sample, the sparse feature model predicts $y=0$ and the dense feature model predicts $y=1$. Recall that the third synthetic dataset is generated from both sparse and dense features, \ie, only when both dense and sparse document features match with the query features, the ground-truth label is $1$. In this case, the \sepattn model mainly focuses on the $0$ predictions as shown in Figure~\ref{fig:attn}(c), which explains why \textbf{\sepattn} works as expected---the final prediction is dominated by negative predictions from either dense or sparse features.

\begin{table*}[h]
  \caption{Evaluations on testing set with percentage of improvements over the best baseline. Metrics with an upper arrow ($\uparrow$) indicate higher is better; metrics with a down arrow ($\downarrow$) indicate lower is better. Improvements with an asterisk (*) denotes statistical significance relative to the best performing baseline (Concatenation), according to the two-tailed paired t-test.}
  \label{tab:eval}
  \scalebox{1.0}{
  \begin{tabular}{*{6}c}
    \toprule
    Methods & MRR ($\uparrow$) & WMRR ($\uparrow$) & ARP ($\downarrow$) & WARP ($\downarrow$) & DCG ($\uparrow$)\\
    \midrule
    Dense-only method & 0.650 & 0.563 & 2.143 & 2.474 & 0.759\\
    Sparse-only method & 0.665 & 0.578 & 2.067 & 2.412 & 0.767\\
    Concatenation method & 0.682 & 0.589 & 2.002 & 2.366 & 0.770\\
    \sepattn & \textbf{0.686} & \textbf{0.595} & \textbf{1.992} & \textbf{2.350} & \textbf{0.781}\\
    \midrule
    $\Delta (\%)$ & $+0.59^*$ & $+1.02^*$ & $-0.50^*$ & $-0.68^*$ & $+1.43^*$\\
  \bottomrule
\end{tabular}
}
\end{table*}

\section{Experiments}
\label{sec:main-exp}
In this section, we begin with a description of the datasets we use in our experiments. Then, we evaluate our proposed technique and baselines using the evaluation metrics that we will define. Finally, we discuss the hyperparameter sensitivity of our method.

\subsection{Dataset}
\label{exp:real-data}
Due to the private and sensitive nature of personal email data, there is no publicly available large-scale email search dataset. Therefore, the data we use comes from the search click logs of Gmail search engine. The training set contains around $317$ million queries, and the testing set contains around $41$ million queries. All queries in the testing set are issued strictly later than all queries in the training set. Each query has six candidate documents (\ie, $n=6$ in Equation~(\ref{eq:func})), one of which is clicked\footnote{These six candidate documents are presented in the dropdown menu of the Gmail search box while users type their queries but before they click the search bottom. When users find the target email, the system will direct them to the exact email, generating exactly one click.}. The goal is to rank the six documents to increase the likelihood of a higher ranked document being clicked.  


\subsection{Model Evaluation}
\label{sec:eval}
In this subsection, we describe the evaluation metrics. 
We denote the evaluation set as $Q$, and the ranking position of the clicked document for query $q$ as $r^*_q$.

\begin{itemize}
\item Mean reciprocal rank (MRR) of the clicked document:
$$
\text{MRR} = \frac{1}{|Q|} \sum_{q\in Q} \frac{1}{r^*_q}.
$$
\item Weighted mean reciprocal rank (WMRR)~\cite{Wang2016LearningTR} of the clicked document:
$$
\text{WMRR} = \frac{1}{\sum_{q\in|Q|} w_q} \sum_{q\in Q} \frac{w_q}{r^*_q},
$$
where $w_q$ is the bias correction weight, and is inversely proportional to the probability of observing a click at the clicked position of $q$. We set those weights using result
randomization, as described in \cite{Wang2016LearningTR}. This serves as our main evaluation metric.
\item Average relevance value position (ARP)~\cite{Zhu2004RecallPA}:
$$
\text{ARP} = \frac{1}{|Q|} \sum_{q\in Q} r^*_q.
$$
\item Weighted average relevance value position (WARP), which is the weighted version of the above ARP metric:
$$
\text{WARP} = \frac{1}{\sum_{q\in|Q|} w_q} \sum_{q\in Q} w_q r^*_q,
$$
where $w_q$ is the bias correction weight as in WMRR.
\item Discounted Cumulative Gain (DCG)~\cite{Jrvelin2002CumulatedGE}:
$$
\text{DCG} = \frac{1}{|Q|} \sum_{q\in Q} \frac{1}{\log_2 (1 + r^*_q)}.
$$
\end{itemize}

\begin{figure*}[h]
  \centering
  \includegraphics[width=\linewidth]{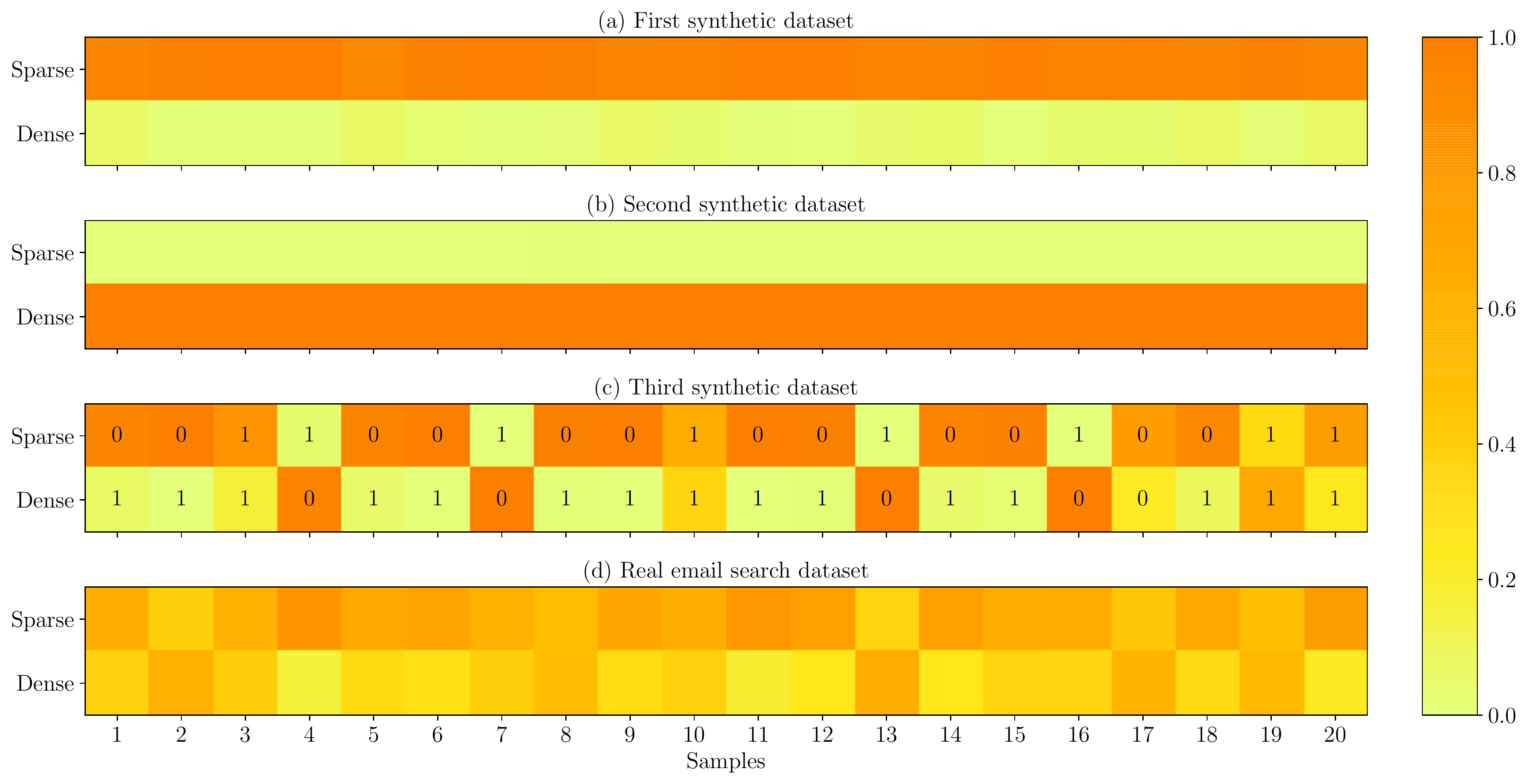}
  \caption{Attention weights visualization ($\alpha_{\text{sparse}}$ and $\alpha_{\text{dense}}$ from Equations~(\ref{eq:alpha_s}) and (\ref{eq:alpha_d})) over $20$ randomly selected samples on four datasets: (a) First synthetic dataset; (b) Second synthetic dataset; (c) Third synthetic dataset; (d) Real email search dataset. The $0$/$1$ values on (c) indicate classification outputs from sparse and dense feature models.}
  \label{fig:attn}
\end{figure*}

\subsection{Results and Discussions}
In this section, we compare our proposed approach, \ie, \textbf{\sepattn} model with the following baseline models. 
\begin{itemize}
\item \textbf{Dense-only method}: The DNN model that only uses dense document features.
\item \textbf{Sparse-only method}: The DNN model that only uses sparse document features.
\item \textbf{Concatenation method}: The DNN model that learns from concatenated dense and embedded sparse features.
\end{itemize}
The above models are implemented using TF-ranking~\cite{Pasumarthi2019TFRankingST} which is a scalable open-source learning-to-rank Tensorflow library~\cite{TF2016}. Due to the data anonymization process (\ie, bag of frequent n-grams), the sequential information in the raw email documents is lost, and thus sequence-aware models like CNNs~\cite{Severyn2015LearningTR} and RNNs~\cite{Pang2017DeepRankAN} cannot be applied to our case. The configuration of the DNN model is described below: The DNN model has three hidden layers whose dimensions are $256$, $128$, and $64$, respectively. The n-grams and character n-grams are first passed through an embedding layer of size $20$ followed by an average pooling layer before fed into the DNN. We use Adagrad~\cite{Duchi2010AdaptiveSM} with $0.1$ learning rate and batch size $100$ to train the model. The above hyperparameters are the optimal settings of the \textbf{Concatenation method}, obtained from tuning the model in the following hyperparameter ranges: Layer dimension in $\{64, 128, 256, 512, 768, 1024\}$; embedding dimension in $\{20, 30, 40, 50, 100\}$; learning rate in $\{0.05, 0.08,
0.1, 0.15, 0.3\}$; optimization algorithm in $\{\text{Adam~\cite{Kingma2014AdamAM}}, \text{Adagrad~\cite{Duchi2010AdaptiveSM}}\}$.

We set the regularization weight to be $1$ in our \textbf{\sepattn} model, and will study model sensitivity to this hyperparameter in Section~\ref{sec:hp}.

The results are presented in Table~\ref{tab:eval}. From the table, we can see that across all metrics, the concatenation of dense features with embedded sparse features (\ie,  Concatenation method) consistently leads to better results than both Sparse-only and Dense-only models. An interesting finding is that even without any sparse features (Dense-only), the model achieves reasonably good performance, probably because emails that are more recent or with more attachments are more likely to be the user-desired ones.

Moreover, our \textbf{\sepattn} method achieves statistically significant improvements (using the two-tailed paired t-test with $99\%$ confidence level) over the Concatenation method.

We note that an improvement of $1\%$ is considered to be highly significant for our email search ranking system. In Table~\ref{tab:eval}, we see that the improvement of WMRR metric for \textbf{\sepattn} method is $1.02\%$ over the best performing baseline model (Concatenation method).


The results demonstrate that dense features indeed play a critical role in email search tasks, and they need to be incorporated into the model effectively, \ie, {simply} concatenating them with embedded sparse features is sub-optimal; \sepattn provides a more effective way of learning from both sparse and dense features.

\subsection{Sensitivity Analysis} 
\label{sec:hp}

\begin{figure}[h]
  \centering
  \includegraphics[width=\linewidth]{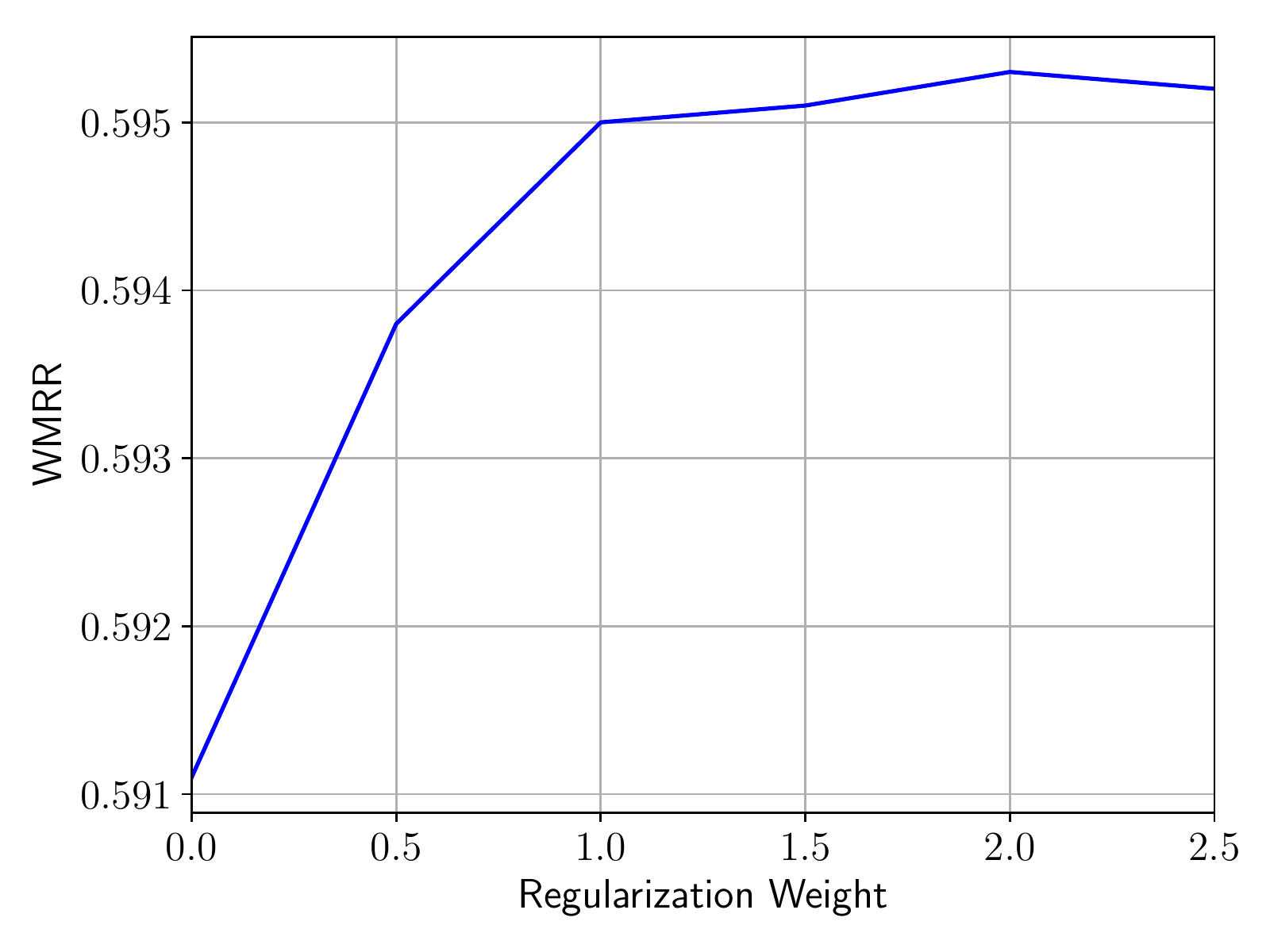}
  \caption{Sensitivity of regularization parameter $\lambda$ on WMRR metric.}
  \label{fig:reg}
\end{figure}

In this section, we discuss the sensitivity of \textbf{\sepattn} method to 
the regularization parameter $\lambda$ in Equation~(\ref{eq:tot}). We vary the regularization parameter $\lambda$ in the range $[0, 2.5]$ and report \sepattn's performance (with respect to WMRR metric) in Figure~\ref{fig:reg}.

We observe that when $\lambda>0$, the performance is significantly better than that when $\lambda=0$. This shows that jointly training the sparse and dense feature models by encouraging them to make consistent prediction is indeed beneficial for improving the ranking performance. When $\lambda$ continues to grow larger, the performance of the model becomes stable, which is a favorable property because $\lambda$ can be safely set within a relatively wide range of values. 

\subsection{Attention Weights Visualization} 
To understand how \textbf{\sepattn} works for different email search queries, we visualize the attention weights ($\alpha_{\text{sparse}}$ and $\alpha_{\text{dense}}$ in Equations~(\ref{eq:alpha_s}) and (\ref{eq:alpha_d})) of \textbf{\sepattn} model on our email search dataset in Figure~\ref{fig:attn}(d). We randomly select $20$ queries. We observe the following: \textbf{(1)} Both sparse and dense features are important for email search tasks, and sparse features generally obtain higher weights than dense features, which corresponds to our intuition that textual features are more important than numerical features in email search. \textbf{(2)} Different queries have different attention weight distribution on sparse and dense features, which verifies the necessity to build an attentive ranking model that learns to focus on different feature sets according to different queries types. For example, the $2$nd, $13$th, $17$th and $19$th samples in Figure~\ref{fig:attn}(d) have higher attention weights on dense features, while for other queries, sparse features are more important. \textbf{\sepattn} automatically learns to assign appropriate weights to dense and sparse features in order to achieve the best performance.

\section{Conclusions and Future Work}
\label{sec:conclude}
In this paper, we studied how to improve email search ranking by effectively learning from a combination of dense and sparse document features using DNNs. We first showed on a set of synthetic datasets that simply concatenating dense features with embedded sparse features leads to sub-optimal performance of DNN ranking models, mainly because these features are from different feature space. Motivated by this drawback, we proposed the \sepattn model which automatically learns to focus on important features for different queries through the \textbf{attention} mechanism. We evaluated our \sepattn model on a large-scale email search dataset and showed that it significantly outperforms the baseline approach---direct concatenation of dense features with sparse features. 

In the future, we would like to extend our study to other IR tasks such as web search and recommendation. Our proposed method can be easily generalized to other scenarios where dense numerical features and sparse categorical features both play a role. Furthermore, the attention mechanism at the prediction level introduced in this paper may facilitate further research on ensembling multiple learning models.

\begin{acks}
We thank Zhongliang Li for his help on the model implementation. We thank anonymous reviewers for valuable and insightful feedback.
\end{acks}

\bibliographystyle{ACM-Reference-Format}
\bibliography{sample-base}

\appendix

\end{document}